\documentclass[conference]{IEEEtran}
\IEEEoverridecommandlockouts
\usepackage{amsmath,graphicx}
\usepackage{amsmath,amssymb,amsfonts}
\usepackage{cite}
\usepackage{url}
\usepackage{bm}
\usepackage{comment}

\usepackage[usenames,dvipsnames]{color}

\title{Deep Joint Source-Channel Coding \\ Over the Relay Channel}
\author{Chenghong Bian, Yulin Shao, Haotian Wu, Deniz G{\"u}nd{\"u}z
\thanks{C. Bian, H. Wu, and D. G{\"u}nd{\"u}z are with the Department of Electrical and Electronic Engineering, Imperial College London (\{c.bian22, haotian.wu17, d.gunduz\}@imperial.ac.uk).
Y. Shao is with the State Key Laboratory of Internet of Things for Smart City, and the Department of Electrical and Computer Engineering, University of Macau, Macau S.A.R. (ylshao@um.edu.mo).
}
\thanks{This work received funding from UKRI for the ERC Consolidator project AIR  (EP/X030806/1)  and the CHIST-ERA project SONATA (CHIST-ERA-20-SICT-004, EP/W035960/1).
}
}

\begin{document}

\newcommand{\figsystem}{
  \begin{figure}
    \centering
    \includegraphics[width=0.8\columnwidth]{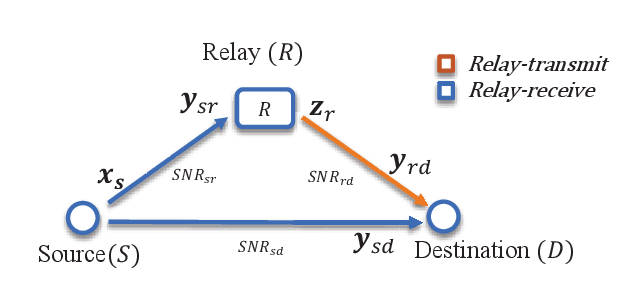}
    \caption{Illustration of the half-duplex relay channel.}
    \label{fig:fig_system}
  \end{figure}
}

\newcommand{\figafdf}{
  \begin{figure}[!t]
    \centering
    \includegraphics[width=0.8\columnwidth]{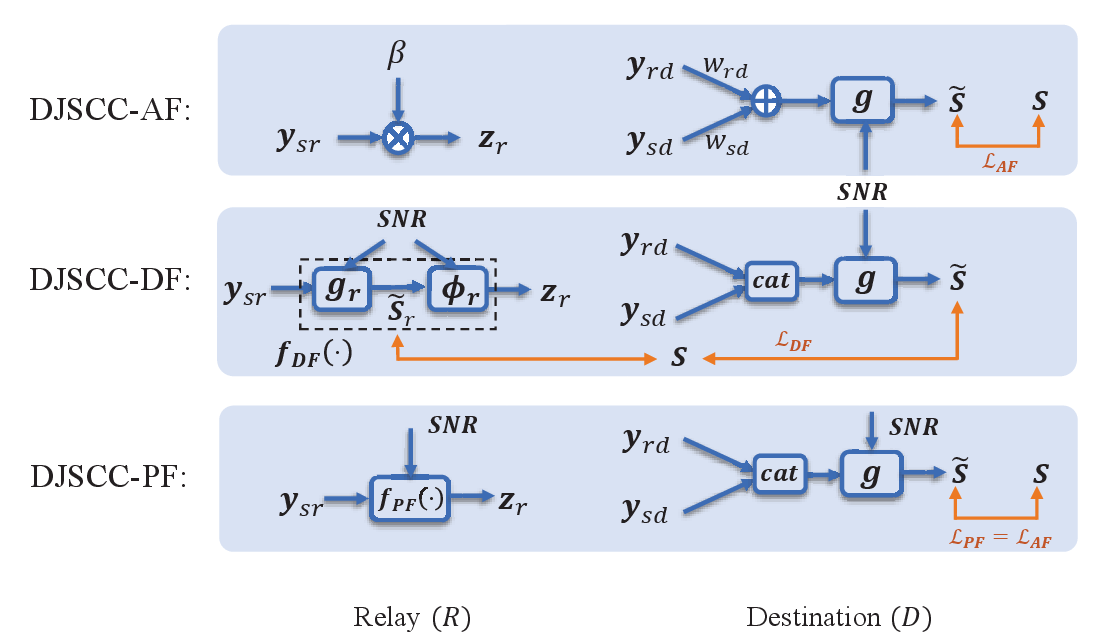}
    \caption{The processing of DeepJSCC-AF, DF and PF at the relay $(\mathrm{R})$ and destination $(\mathrm{D})$, where $\beta$ is the scaling factor while $w_{rd}$ and $w_{sd}$ are the MRC coefficients in \eqref{equ:MRC}. $\bm{SNR}$ denotes the collection of channel qualities, consisting of $SNR_{sr}, SNR_{rd}$ and $SNR_{sd}$.}
    \label{fig:fig_afdf}
  \end{figure}
}

\newcommand{\figneuralnet}{
  \begin{figure}[t]
    \begin{minipage}[b]{\linewidth}
       \centering
       \centerline{\includegraphics[width=7 cm]{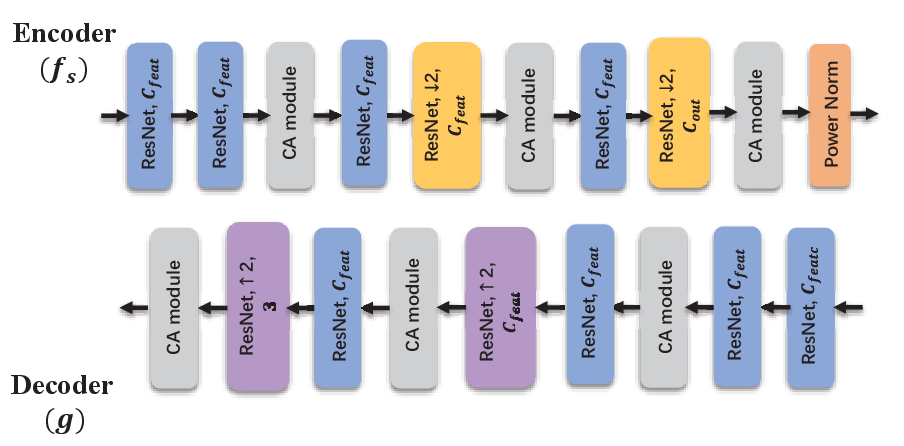}}
       \centerline{{(a)}}\medskip
    \end{minipage}
    \hfill
    \begin{minipage}[b]{\linewidth}
       \centering
       \centerline{\includegraphics[width=6.5 cm]{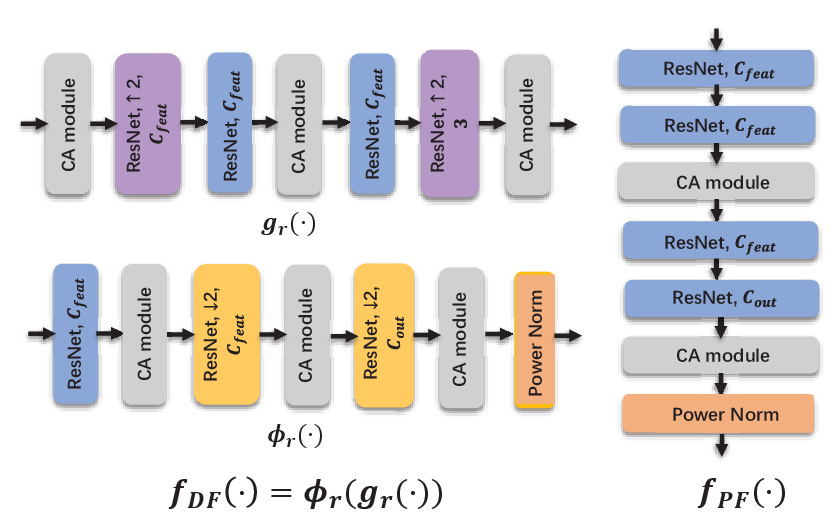}}
       \centerline{{(b)}}\medskip
    \end{minipage}
\caption{{The architectures of the DNNs used to parameterize $f_s, g$ and $f_{PF}, f_{DF}$ functions are shown in (a) and (b), respectively.}}
\label{fig:fig_NN}
  \end{figure}
}

\newcommand{\figAdapt}{
  \begin{figure}[!t]
    \centering
    \includegraphics[width=0.8\columnwidth]{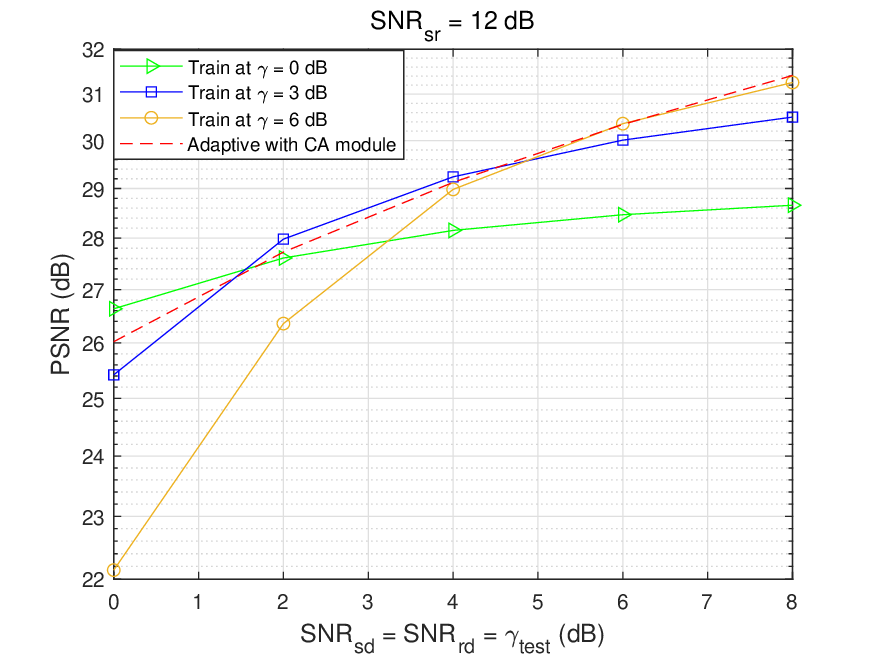}
    \caption{DeepJSCC-PF model with CA modules trained at varying SNR $(\gamma)$ values is compared to the models trained at a fixed $\gamma$ when tested at different $\gamma_{test}$. We set $SNR_{sr} = 12$ dB.}
    \label{fig:fig_SA}
  \end{figure}
}

\newcommand{\figcompafdf}{
  \begin{figure}[!t]
    \centering
    \includegraphics[width=0.8\columnwidth]{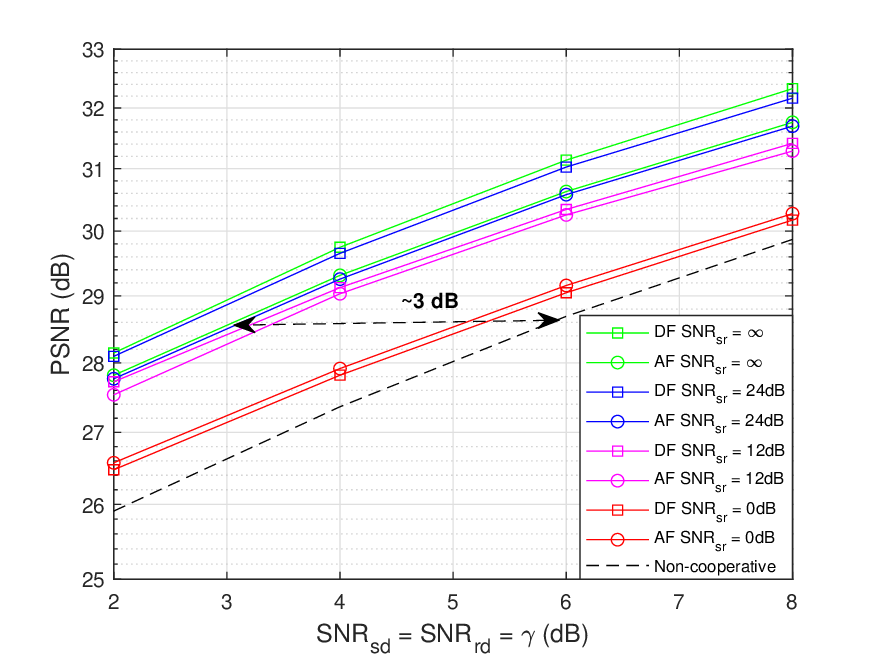}
    \caption{Comparison between DeepJSCC-AF and DeepJSCC-DF with $SNR_{sr} \in \{0, 12, 24, \infty\}$ dB. We also include the non-cooperative scheme \cite{deepjscc} as a benchmark.}
    \label{fig:fig_compare_afdf}
  \end{figure}
}

\newcommand{\figfinal}{
  \begin{figure}[t]
    \begin{minipage}[b]{\linewidth}
       \centering
       \centerline{\includegraphics[width=7 cm]{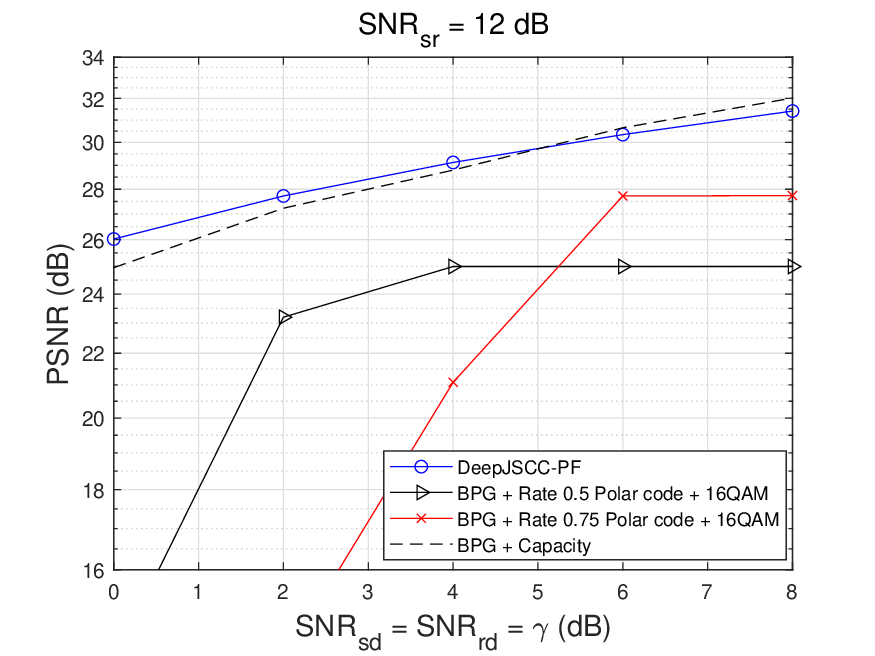}}
       \centerline{(a) PSNR}\medskip
    \end{minipage}
    \hfill
    \begin{minipage}[b]{\linewidth}
       \centering
       \centerline{\includegraphics[width=7 cm]{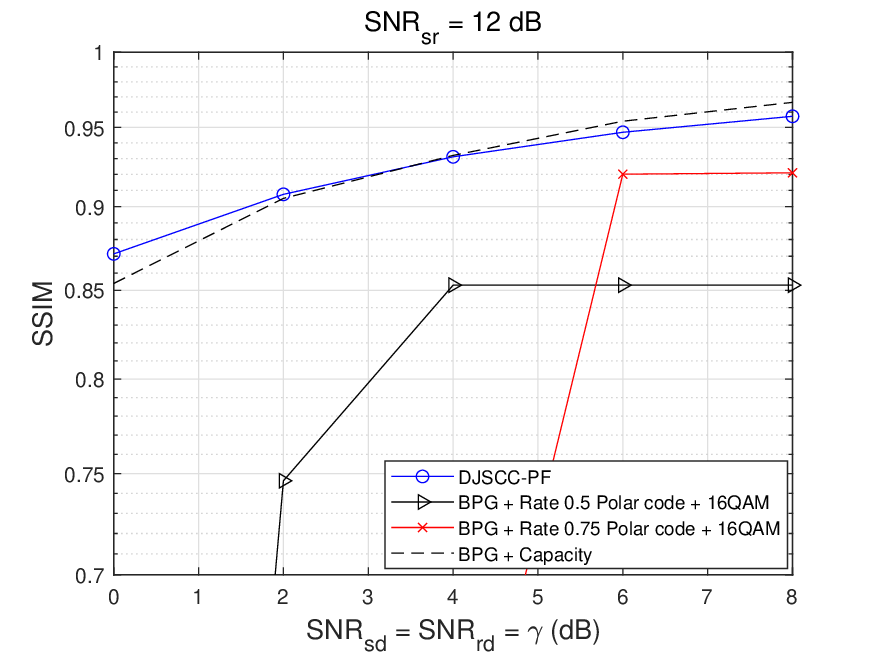}}
       \centerline{(b) SSIM}\medskip
    \end{minipage}
\caption{The PSNR and SSIM performance of the DeepJSCC-PF scheme compared with separation based schemes employing BPG algorithm using either Polar codes, or assuming capacity {achieving channel codes for} $SNR_{sr} = 12$ dB.}
\label{fig:fig_final_results}
  \end{figure}
}

\newcommand{\mytable}{
\begin{table}[tbp]
\caption{{Evaluation for DeepJSCC-DF and DeepJSCC-PF (Top: PSNR, Bottom: SSIM)}}
\begin{center}
\begin{tabular}{c|cccc|c}
\hline
\textbf{Protocol}&\multicolumn{4}{|c|}{\textbf{DF}} &\multicolumn{1}{|c}{\textbf{PF}} \\
\cline{1-6} 
\textbf{$\lambda$} & \textbf{0}& \textbf{0.5}& \textbf{1} & \textbf{2}& \textbf{None}\\
\hline
$SNR_{sr}=0$ dB & \textbf{30.196}& 29.398& 29.157& 28.825 & 30.176\\
$SNR_{sr}=8$ dB & 30.887& \textbf{30.988}& 30.926& 30.884 & 30.910\\
$SNR_{sr}=24$ dB & 32.168& 32.137& \textbf{32.241}& 32.083 & 32.164 \\
\hline
\end{tabular}
\label{tab1}
\end{center}

\vspace{0.2cm}
\begin{center}
\begin{tabular}{c|cccc|c}
\hline
\textbf{Protocol}&\multicolumn{4}{|c|}{\textbf{DF}} &\multicolumn{1}{|c}{\textbf{PF}} \\
\cline{1-6} 
\textbf{$\lambda$} & \textbf{0}& \textbf{0.5}& \textbf{1} & \textbf{2}& \textbf{None}\\
\hline
$SNR_{sr}=0$ dB & \textbf{0.9447}& 0.9310& 0.9255& 0.9194 & 0.9444\\
$SNR_{sr}=8$ dB & 0.9517& \textbf{0.9524}& 0.9519& 0.9518 & 0.9522\\
$SNR_{sr}=24$ dB & 0.9637& 0.9635& \textbf{0.9640}& 0.9632 & 0.9635\\
\hline
\end{tabular}
\label{tab2}
\end{center}
\end{table}
}


\newcommand{\figneuralnetappendix}{
  \begin{figure}[h]
    \begin{minipage}[b]{\linewidth}
       \centering
       \centerline{\includegraphics[width=7 cm]{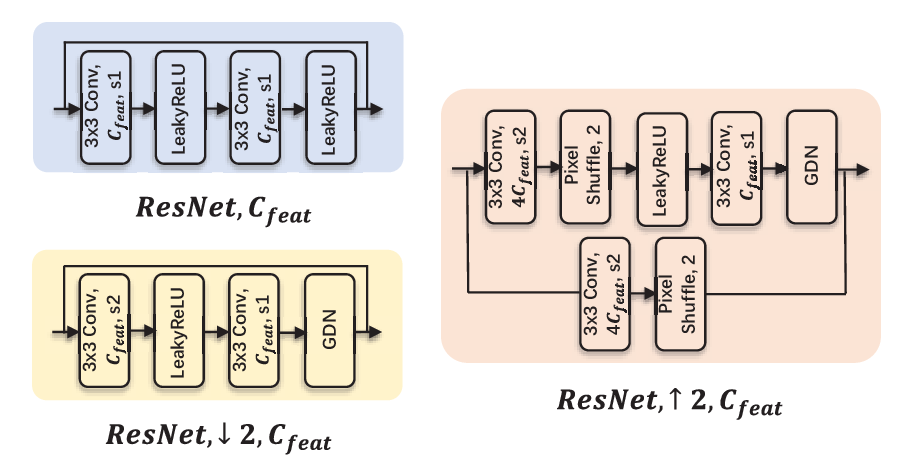}}
       \centerline{{(a)}}\medskip
    \end{minipage}
    \hfill
    \begin{minipage}[b]{\linewidth}
       \centering
       \centerline{\includegraphics[width=8 cm, height = 2.5cm]{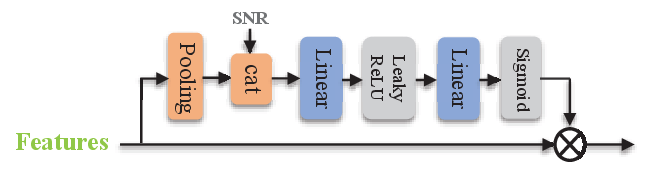}}
       \centerline{{(b)}}\medskip
    \end{minipage}
\caption{The neural network architectures of `ResNet', `ResNet, $\uparrow$ 2', `ResNet $\downarrow$ 2' are shown in (a) while the CA module is shown in (b).}
\label{fig:fig_NN_appendix}
  \end{figure}
}

\newcommand{\figadaptappendix}{
  \begin{figure}[t]
    \centering
    \includegraphics[width=0.8\columnwidth]{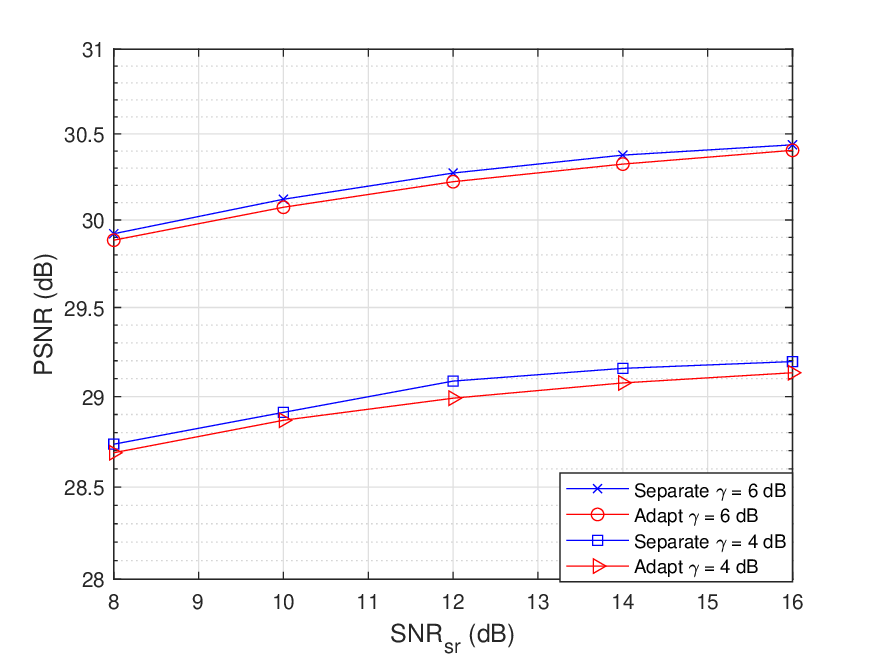}
    \caption{Fully-adaptive DeepJSCC-PF model trained with varying $SNR_{sr}$ values is compared to the models that are trained for a fixed $SNR_{sr}$.}
    \label{fig:fig_adapt_appendix}
  \end{figure}
}

\newcommand{\figcliffeffect}{
  \begin{figure}[t]
    \centering
    \includegraphics[width=0.8\columnwidth]{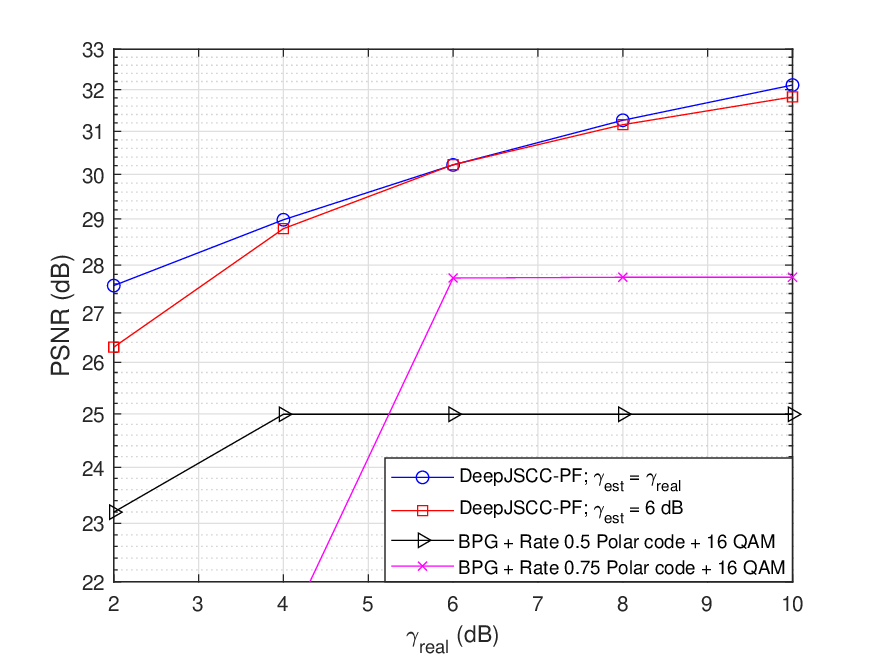}
    \caption{The PSNR performance of the DeepJSCC-PF model with a mismatch between the estimated channel ($\gamma_{est}$) and actual channel quality($\gamma_{real}$). We set $\gamma_{est} = 6$ dB, and test for $\gamma_{real} \in \{2, 4, 6, 8, 10\}$ dB. }
    \label{fig:fig_cliff_effect}
  \end{figure}
}
%

\maketitle
\begin{abstract}
This paper presents a novel deep joint source-channel coding (DeepJSCC) scheme for image transmission over a half-duplex cooperative relay channel. Specifically, we apply DeepJSCC to two basic modes of cooperative communications, namely amplify-and-forward (AF) and decode-and-forward (DF). In DeepJSCC-AF, the relay simply amplifies and forwards its received signal. In DeepJSCC-DF, on the other hand, the relay first reconstructs the transmitted image and then re-encodes it before forwarding. Considering the excessive computation overhead of DeepJSCC-DF for recovering the image at the relay, we propose an alternative scheme, called DeepJSCC-PF, in which the relay processes and forwards its received signal without necessarily recovering the image.
Simulation results show that the proposed DeepJSCC-AF, DF, and PF schemes are superior to the digital baselines with BPG compression with polar codes and provide a graceful performance degradation with deteriorating channel quality. Further investigation shows that the PSNR gain of DeepJSCC-DF/PF over DeepJSCC-AF improves as the channel condition between the source and relay improves. Moreover, the DeepJSCC-PF scheme achieves similar performance to DeepJSCC-DF with lower computational complexity.
\end{abstract}
%


\section{Introduction}
\label{sec:intro}

The relay channel consists of three terminals, the source (S), the relay (R), and the destination (D) (see Fig. \ref{fig:fig_system} for an illustration). The source broadcasts its message to the relay and destination; then, the relay processes its received signal and forwards it to the destination, while the destination performs joint decoding to recover the original message. There are three classical relaying protocols: amplify-and-forward (AF), decode-and-forward (DF) and compress-and-forward (CF) \cite{relay_capacity, relay_capacity2, CF}, and their variations \cite{orth_relay, deter_relay}. In AF, the relay simply scales its received signal, and forwards to the destination. The main limitation of AF is the inherent noise forwarding. In DF, the relay decodes the received signal into the original bit sequence, and then re-encodes. While DF mitigates the noise forwarding problem, its performance becomes limited when the source-to-relay channel is poor. In CF, the relay forwards its received signal to the destination, but unlike in AF, it compresses the received signal using Wyner-Ziv source coding treating the signal received at the destination as side information. 

Shannon's separation theorem applies to the relay channel  (when there is no side information at the relay) \cite{titjscc_relay}; that is, separate compression followed by cooperative channel coding is optimal in the infinitely long source and channel blocks. Interestingly, separation theorem holds even though we do not know the capacity of the relay channel. On the other hand, there are limited studies on joint source-channel coding (JSCC) over relay channels. JSCC over cooperative relay networks is studied in \cite{titjscc_relay} from an information theoretic perspective. JSCC for cooperative transmission of multimedia sources have been considered in \cite{coop_multimedia}; however, these papers consider separate codes for compression and error correction, whose parameters are optimized jointly to create robustness against channel variations. 

\figsystem

Recently, deep neural networks (DNNs) have been successfully applied to a wide range of communication problems, e.g., channel coding \cite{sparc} and signal processing \cite{mimo_det}. Due to its strong capability to extract important features, DNNs have also been successfully applied to the JSCC problems, and the resultant scheme is shown to outperform digital alternatives employing {state-of-the-art compression techniques together with near-optimal channel codes}, as well as providing graceful degradation with degrading channel quality \cite{deepjscc}. {The DeepJSCC technique} introduced in \cite{deepjscc} is further extended to different sources such as text, video and point clouds \cite{deepjscc_text, semantic_video, bian2023wireless}. Recent works show that DeepJSCC can also adapt to more challenging multipath fading channel\cite{jsccofdm}, multi-input multi-output (MIMO) channel\cite{st_jscc}  and multihop relay channel\cite{semantic_multihop1}. To be best of our knowledge, there is no prior work that studies DeepJSCC in the cooperative relay setting, which, as we will show in this paper, is a non-trivial extension of the prior work, {and is an important step towards realizing JSCC in practical multiuser networks.}

In this paper, we propose three different DeepJSCC schemes, namely DeepJSCC-AF, DF, and PF for image transmission over a cooperative relay channel.
DeepJSCC-AF and DeepJSCC-DF schemes resemble the classical AF and DF protocols in relay networks, while the DeepJSCC-PF is a variation of DeepJSCC-DF with a simplified neural network at the relay for reduced complexity. 
To be specific, in DeepJSCC-AF, the relay amplifies its received signal while the destination node applies the maximum ratio combining (MRC) rule to combine the two received signals. In DeepJSCC-DF, the DNN at the relay reconstructs the original image, and re-encodes it to generate additional `parity' symbols to be forwarded to the destination, which concatenates the signals from the source (termed as `systematic' symbols) with those from the relay to decode the image. 
DeepJSCC-PF is similar to DeepJSCC-DF, except that it does not require the relay to explicitly reconstruct the image, which reduces both the computation load and the delay at the relay. 
Our numerical results show that all the proposed schemes outperform the digital baseline with BPG compression followed by polar-coded channel transmission while avoiding the cliff effect.  Finally, we reveal the robustness of DeepJSCC-DF/PF against poor source-to-relay channel condition and their superior performance over DeepJSCC-AF {when the source-to-relay channel is strong}.
\vspace{-0.3cm}

\figneuralnet

\section{System model}\label{sec:system_model}
As illustrated in Fig.~\ref{fig:fig_system}, we consider a classical relay channel model consisting of a source node $\mathrm{S}$, a destination node $\mathrm{D}$, and a relay node $\mathrm{R}$.  The goal is to deliver an image $\bm{S} \in \mathbb{R}^{C\times H \times W}$ from $\mathrm{S}$ to $\mathrm{D}$ with the help of relay $\mathrm{R}$, where $C$, $H$, $W$ denote the number of color channels, the height and width of the image, respectively. We assume a half-duplex relay node that cannot receive and transmit at the same time. Therefore, the transmission is divided into two periods \cite{relay_capacity}: the \textit{relay-receive} period and the \textit{relay-transmit} period. In the former, $\mathrm{S}$ encodes the image $\bm{S}$ into a channel codeword $\bm{x}_s \in \mathbb{C}^k$ using an encoder function $f_s: \mathbb{R}^{C\times H \times W} \rightarrow \mathbb{C}^k$, where $\bm{x}_s = f_s(\bm{S})$. The source input is subject to a power constraint:
\begin{equation}
\frac{1}{k} \|\bm{x}_s \|^2 \leq 1.
\end{equation}
In the context of JSCC, the `bandwidth ratio' quantifies the available channel uses per pixel (CPP), defined as $\rho \triangleq \frac{k}{C \cdot H \cdot W}$.

The received signal at the relay $\mathrm{R}$ and destination $\mathrm{D}$ at the \textit{relay-receive} period can be written as
\begin{equation}
\label{equ:relay-receive}
    \bm{y}_{sr} = \alpha_{sr}\bm{x}_s + \bm{n}_{r},
\end{equation}
\begin{equation}
\label{equ:destination-receive}
    \bm{y}_{sd} = \alpha_{sd}\bm{x}_s + \bm{n}_{d},
\end{equation}
where $\alpha_{sr}$, $\alpha_{sd}$ are real constants governed by the transmission distances of the $\mathrm{S}-\mathrm{R}$ and $\mathrm{S}-\mathrm{D}$ links, respectively; $\bm{n}_{r} \sim \mathcal{CN}(0,N_r), \bm{n}_{d} \sim \mathcal{CN}(0,N_d)$ denote complex additive white Gaussian noise (AWGN) signals. The {signal-to-noise ratio (SNR)} of the $\mathrm{S}-\mathrm{R}$ and $\mathrm{S}-\mathrm{D}$ links are defined as
\begin{equation}
SNR_{sr} \triangleq \frac{\alpha_{sr}^2}{N_r}~~~ \mbox{and   }
    SNR_{sd} \triangleq \frac{\alpha_{sd}^2}{N_d}.
    \label{equ:SNR}
\end{equation}

Upon receiving $\bm{y}_{sr}$, the relay processes it by a function $f_r(\cdot)$ and forwards the processed signal to the destination in the relay-transmit period. Denoting the signal transmitted from the relay by $\bm{z}_r \in \mathbb{C}^k$, we have
\begin{equation}
\bm{z}_r= f_r(\bm{y}_{sr}),
\end{equation}
where $\bm{z}_r$ is also subject to a power constraint:
$\frac{1}{k} \|\bm{z}_r\|^2 \leq 1$. 

The signal received at the destination in the \textit{relay-transmit} period can be written as:
\begin{equation}
    \bm{y}_{rd} = \alpha_{rd}\bm{z}_r + \bm{n}_{rd},
    \label{equ:relay-transmit}
\end{equation}
where $\alpha_{rd}, \bm{n}_{rd}$ and $SNR_{rd}$ are defined similarly as in \eqref{equ:destination-receive} and \eqref{equ:SNR}.

Given the received signal from the source and relay in the two periods, the destination reconstructs the image using a decoding function $g = \mathbb{C}^k \times \mathbb{C}^k \rightarrow \mathbb{R}^{C\times H \times W}$. The reconstructed image is denoted by $\widetilde{\bm{S}} \in \mathbb{R}^{C\times H \times W}$, where
\begin{equation}
\widetilde{\bm{S}} = g(\bm{y}_{sd}, \bm{y}_{rd}).
\end{equation}
The peak signal-to-noise ratio (PSNR) and the structural similarity index measure (SSIM) will be used to evaluate the reconstruction quality of $\widetilde{\bm{S}}$.

We remark here that we do not allow the source to continue transmission during the relay-transmit period. More general schemes that go beyond these limitations will be considered in future work.

\section{DeepJSCC for Relay-Aided Communications}\label{sec:AFDF}
We parameterize the encoder $f_s$, decoder $g$, and the transformation {$f_r$} at the relay by DNNs. {Function $f_r$ at the relay terminal depends on the particular relaying scheme. In the case of DeepJSCC-AF, the relay simply amplifies its received signal, and no DNN is needed. The architectures used for the relay function in DeepJSCC-DF and DeepJSCC-PF schemes, denoted by $f_{DF}$ and $f_{PF}$, respectively, are shown in Fig. \ref{fig:fig_NN}(b).} 
Specifically, `ResNet' in Fig. \ref{fig:fig_NN} refers to a block of 2D convolutional neural networks with residual connections. We emphasize that residual connections are essential in our problem, especially at the relay. 
The up-sampling blocks in $g$ are realized by pixel shuffling. The parameter $C_{fea t}$ denotes the number of filters and $C_{out}$ is determined by the bandwidth ratio $\rho$.

Inspired by \cite{xu2021wireless}, we introduce the channel attention (CA) module to improve the robustness to different channel conditions. These CA modules take both the features and channel conditions as inputs, where the channel conditions refer to a collection of SNRs defined as $\bm{SNR} = (SNR_{sr}, SNR_{sd}, SNR_{rd})$, allowing DNNs to assign different weights to different input features according to the channel conditions. Based on the DNNs described above, we next explain the processing at the relay and destination nodes when operating under DeepJSCC-AF, DF and PF protocols, respectively.

\figafdf

\subsection{DeepJSCC-AF}
As shown in Fig.~\ref{fig:fig_afdf}, {$f_{AF}$} is simply a linear function, and the signal transmitted by the relay is a scaled version of its received signal: is,
\begin{equation}
 \bm{z}_r = \beta \bm{y}_{sr},
\end{equation}
where $\beta \triangleq \sqrt{\frac{1}{\alpha_{sr}^2+N_r}}$ ensures that the power constraint is satisfied. Substituting $\bm{z}_r$ into \eqref{equ:relay-transmit} yields
\begin{equation}
    \bm{y}_{rd} = \beta \alpha_{rd}\alpha_{sr}\bm{z} + \widetilde{\bm{n}},
    \label{equ:AF}
\end{equation}
where $\widetilde{\bm{n}}$ is AWGN with zero mean and variance $N_d+\frac{N_r \alpha_{rd}^2}{\alpha_{sr}^2+N_r}$. 

Given \eqref{equ:destination-receive} and \eqref{equ:AF}, the received signal at the destination from two transmission periods can be combined by MRC to generate an estimate of $\bm{z}$, expressed as:
\begin{equation}
    \widetilde{\bm{z}} = \frac{(\beta^2\alpha_{rd}^2 N_r + N_d)\alpha_{sd}\bm{y}_{sd} + N_d \beta \alpha_{rd}\alpha_{sr}\bm{y}_{rd}}{N_d \beta^2 \alpha_{rd}^2 \alpha_{sr}^2+ \alpha_{sd}^2(\beta^2 \alpha_{rd}^2 N_r + N_d)}.
    \label{equ:MRC}
\end{equation}
Then we convert $\widetilde{\bm{z}}$ into a real vector, reshape it, and feed it to the DeepJSCC-AF decoder along with $\mathbf{SNR}$. The whole system is trained in an end-to-end fashion with mean square error (MSE) as the loss function: 
\begin{equation}
    \mathcal{L}_{AF} = \mathbb{E}_{\bm{S}}\left[||\bm{S} - \widetilde{\bm{S}}||^2_2\right].
    \label{equ:Loss_AF}
\end{equation}

\subsection{DeepJSCC-DF}
In the conventional (digital) DF protocol, the relay decodes the original information bits, re-encodes the bits to create additional parity symbols, and forwards these parity symbols to the destination. We can apply a similar technique in conjunction with DeepJSCC: the relay tries to reconstruct the source image from $\bm{y}_{sr}$, then performs another DeepJSCC encoding on the recovered image into $\bm{z}_r$ for transmission. Specifically, we first feed $\bm{y}_{sr}$ to the relay decoder network $g_r$ (along with the {$\bm{SNR}$}), yielding an estimated image $\widetilde{\bm{S}}_r = g_r(\bm{y}_{sr}, \bm{SNR})$ followed by the relay encoder $\phi_r(\cdot)$ to generate $\bm{z}_r$:
\begin{align}
\label{equ:DeepJSCC-DF}
    \bm{z}_r &= \phi_r(\widetilde{\bm{S}}_r, \bm{SNR}).
\end{align}
Note that $\bm{z}_r$ will be power normalized before transmission. As shown in Fig.~\ref{fig:fig_afdf}, we use $f_{DF} = \phi_r(g_r(\cdot))$ to represent the operations at the relay in the DF mode.
At the destination, we concatenate the received signals $\bm{y}_{sd}$ and $\bm{y}_{rd}$ to reconstruct $\tilde{\bm{S}}$.

The loss function of the proposed DeepJSCC-DF is comprised of two parts, representing the image reconstruction quality at the relay and the destination, respectively.
To balance the two, we introduce a hyper-parameter $\lambda \ge 0$, and the overall loss function is given by
\begin{equation}
    \mathcal{L}_{DF} = \mathbb{E}_{\bm{S}}\left[||\bm{S} - \widetilde{\bm{S} }||^2_2 + \lambda ||\bm{S} - \widetilde{\bm{S}}_r ||^2_2\right].
    \label{equ:DFLoss}
\end{equation}
Intuitively, a smaller $\lambda$ emphasizes the reconstruction quality at the destination. {We note that, unlike the DF scheme which is based purely on channel coding, DeepJSCC-DF cannot prevent error propagation, since the reconstruction at the relay is always prone to some reconstruction errors.}

\subsection{DeepJSCC-PF}
{Although the DeepJSCC-DF does not require a high-fidelity reconstruction of the source image at the relay, when $\lambda = 0$, its architecture is designed assuming the source will be recovered to generate the parity symbols.} This results in non-negligible computing-resource consumption and higher delays compared to the DeepJSCC-AF scheme. In this context, we introduce an alternative relaying protocol called DeepJSCC-PF that achieves comparable reconstruction results with DeepJSCC-DF. In particular, DeepJSCC-PF does not require explicit reconstruction at the relay, employing a lightweight neural network to model relay processing.


We have $\bm{z}_r = f_{PF}(\bm{y}_{sr})$, which is normalized before transmission. The decoder for DeepJSCC-PF at the destination is identical to that of DeepJSCC-DF: the received signals are concatenated, reshaped, and fed into the decoder. The training loss $\mathcal{L}_{PF}$ is the same as that of DeepJSCC-AF. The relaying and decoding processes for DeepJSCC-AF, DF, and PF are illustrated in Fig.~\ref{fig:fig_afdf}.

\section{Numerical Experiments}

\label{sec:experiment}
\subsection{Parameter Settings and Training Details}
Throughout this section, we assume that the $\mathrm{S}-\mathrm{D}$ and $\mathrm{R}-\mathrm{D}$ links have the same quality, while the $\mathrm{S}-\mathrm{R}$ link has better quality, i.e., $SNR_{sr}>SNR_{sd}=SNR_{rd} \triangleq \gamma$.

We evaluate the proposed DeepJSCC-AF, DF, and PF architectures considering the transmission of images from the CIFAR-10 dataset, which consists of $50,000$ training and $10,000$ test RGB images with $32 \times 32$ resolution. The Adam optimizer is adopted to train the model with a varying learning rate. Specifically, the learning rate is initialized to $10^{-4}$ and will be dropped by a factor of $0.8$ if the validation loss does not improve in $4$ consecutive training epochs.
Throughout this section, we fix $\rho = 0.125$, and hence, $C_{out} = 12$.

\subsection{Performance evaluation}

\figAdapt

\subsubsection{{Robustness to test SNR variations}}
Wefirst evaluate the robustness of the DeepJSCC architecture with the CA module to test SNR variations.
The performance of the proposed DeepJSCC-PF scheme\footnote{We note that DeepJSCC-AF and DF with CA modules are also robust to test SNR variations, but we provide the results only for PF due to page limitations.} trained with varying SNR values $\gamma \in [2,8]$ dB is shown in Fig.~\ref{fig:fig_SA}.
In both training and testing, we set $SNR_{sr} = 12$ dB. We then consider training the networks for fixed values of $SNR_{sd}$ = $SNR_{rd} = \gamma \in \{0, 3, 6\}$ dB without the CA module. In the test phase, these models are evaluated at $\gamma_{test} \in [0,8]$ dB.

As can be seen from Fig.~\ref{fig:fig_SA}, with the CA module  (i.e., the curve labeled with `adaptive with CA module'), the well-trained DeepJSCC model is robust to SNR variations in the test phase. Under a given test SNR, the PSNR performance achieved by the model with the CA module is comparable with the one trained under the given test SNR. We adopt CA modules in all the following simulations.

\subsubsection{{Comparison of different relaying protocols}} 
Next, we compare the performances of DeepJSCC-AF and DeepJSCC-DF, benchmarked against non-cooperative transmission, which refers to transmission from $\mathrm{S}$ to $\mathrm{D}$ in the absence of $\mathrm{R}$ \cite{deepjscc}. The non-cooperative scheme uses the same neural network structure while the source utilizes $k$ symbols to transmit the image.
The PSNR comparisons of these schemes are shown in Fig.~\ref{fig:fig_compare_afdf}, where we train and test different DeepJSCC-AF and DF pairs with $SNR_{sr}\in\{0, 12, 24,\infty\}$ dB. Note that DeepJSCC-DF curves shown in this simulation are obtained with the optimal $\lambda$ for each $SNR_{sr}$ value, which will be detailed later.

\figcompafdf

In {cooperative channel coding}, the condition of the $\mathrm{S}-\mathrm{R}$ link determines the relative performances of AF, DF, and non-cooperative transmission {in terms of the achievable rates} \cite{relay_capacity2}.
In particular, when $SNR_{sr}$ is low, it is no longer possible to decode at the relay; and hence, DF becomes worse than AF and non-cooperative transmission in this regime.
Our results in Fig.~\ref{fig:fig_compare_afdf} tell a different story: the PSNR of our DeepJSCC-DF scheme is strictly better than that of the non-cooperative transmission, even when $SNR_{sr} = 0$ dB. This is because, even though we require the relay to decode the image, we allow lossy reconstruction in DeepJSCC. Therefore, in DeepJSCC-DF, the reconstruction quality at the destination is not limited by that of the relay, unlike in the digital DF scheme.

{The performance gain of DeepJSCC-DF with respect to DeepJSCC-AF} increases with $SNR_{sr}$. In particular, when $SNR_{sr} = \infty$, DeepJSCC-DF outperforms DeepJSCC-AF by approximately $1$ dB. This can be understood as a coding gain: when $SNR_{sr} = \infty$, DeepJSCC-AF is essentially a repetition code. Notice that we have set $SNR_{sd} = SNR_{rd}$, as a result, the PSNR performance of DeepJSCC-AF is $3$ dB better than that of the non-cooperative transmission. 
For DeepJSCC-DF, the relay generates `parity' symbols from the received signal, providing an additional coding gain compared to DeepJSCC-AF. 
However, under lower $SNR_{sr}$, the noise in $\bm{y}_{sr}$ will hinder the parity generation process and the coding gain vanishes.

We then study the effect of the hyper-parameter $\lambda$ introduced in Section. \ref{sec:AFDF} and compare the DeepJSCC-PF and DeepJSCC-DF schemes in Table \ref{tab1}. In the training phase, different DeepJSCC-DF and PF models are trained with $SNR_{sr} \in \{0, 8, 24\}$ dB, $\gamma \in \mathcal{U}(2,8)$ dB and we consider $\lambda \in \{0,0.5,1,2\}$ for the DeepJSCC-DF models. In the testing phase, $\gamma = 8$ dB is set for all the simulations\footnote{We further note that the conclusions reached for  $\gamma = 8$ dB hold for other $\gamma$ values too. These results are not included here due to space constraints.}.
\mytable
As one would expect, when $SNR_{sr} = 0$ dB, $\lambda = 0$ is preferable as trying to reconstruct the image at the relay {becomes a bottleneck due to poor $\mathrm{S}-\mathrm{R}$ link quality. When {$SNR_{sr} = 8, 24$ dB}, a non-zero $\lambda$ ensures a good reconstruction at the relay, preventing noise forwarding to some extend. We note that, even in this case, increasing $\lambda$ beyond 1 harms the final performance as the system is mainly focusing on the reconstruction at the relay. 

From Table \ref{tab1}, we observe that the gain obtained by tuning $\lambda$ is limited: DeepJSCC-DF with $\lambda = 0$ performs close to the {the results with optimal $\lambda$ for each SNR value}. This motivates the introduction of the DeepJSCC-PF scheme, which is equivalent to DeepJSCC-DF with $\lambda = 0$, with a simplified DNN at the relay. We observe from Table I that DeepJSCC-PF can achieve comparable performance with DeepJSCC-DF with $\lambda = 0$ using a much simpler DNN architecture.

\figfinal

\subsubsection{{Comparison with the digital baseline}} 
Finally, we compare the proposed DeepJSCC-PF scheme with a digital baseline in Fig.~\ref{fig:fig_final_results}. In this simulation, we keep $SNR_{sr} = 12$ dB and vary $\gamma$ from $0$ dB to $8$ dB. For the baseline, we use the state-of-the-art BPG compression algorithm with the 3GPP polar code. To exploit the full potential of the baseline, we assume that the relay can always decode $\bm{y}_{sr}$ perfectly.
The processing of the digital baseline is as follows: the source node first compresses the image via BPG and encodes it using a rate $R\in (0,1)$ polar code followed by a given modulation scheme. The relay perfectly decodes the information bits, generates parity bits, and forwards them to the destination in the second half. The received signal at the destination is a noisy version of the rate $R/2$ polar code with the same modulation. We also provide an upper bound on the performance of the separation scheme by calculating the capacity of the system described above (with perfect $\mathrm{S}-\mathrm{R}$ link), given by $\log_2(1+10^{\gamma/10})$. It can be seen that the proposed DeepJSCC-PF outperforms the separation-based approach at all noise levels in terms of both PSNR and SSIM. We note that DeepJSCC-PF has comparable performance even with the ideal separation-based scheme with capacity-achieving channel codes, which also assumes an ideal $\mathrm{S}-\mathrm{R}$ link. 

\figcliffeffect

It is well known that the key advantage of DeepJSCC over the conventional schemes is that the neural network is able to provide graceful degradation when the channel quality changes from the estimated one. In the next simulation, we adopt the aforementioned DeepJSCC-PF model with the same training settings. In the evaluation phase, we set $SNR_{sr} = 12$ dB and assume the estimated channel quality for $\mathrm{S}-\mathrm{D}$ and $\mathrm{R}-\mathrm{D}$ links are $\gamma_{est} = 6$ dB. The actual channel quality $\gamma_{real}$, however, is set to $\gamma_{real} \in \{2, 4, 6, 8, 10\}$ dB. As shown in Fig. \ref{fig:fig_cliff_effect}, we compare the PSNR performance of the `DeepJSCC-PF, $\gamma_{est} = 6$ dB' curve where the channel quality estimation error exists with 
the `DeepJSCC-PF, $\gamma_{est} = \gamma_{real}$' curve where the estimated channel quality always matches the actual quality. The two curves have the same performance at $\gamma_{real} = 6$ dB; however, when channel estimation error occurs, the proposed scheme provides graceful degradation. On the other hand, the digital baseline fails when $\gamma_{real}$ is below a certain threshold.

\subsubsection{Robustness to S-R Channel Conditions} 
We extend the results shown in Fig. \ref{fig:fig_SA} and illustrate that a fully-adaptive DeepJSCC-PF model can also adapt to different source-to-relay channel conditions. In this experiment, the fully-adaptive model is trained with $SNR_{sr} \in [6, 18]$ dB and $\gamma \in [2, 10]$ dB. We then consider different networks which are only adaptive to $\gamma$. In particular, five models are trained under fixed $SNR_{sr} \in \{8, 10, 12, 14, 16\}$ dB with $\gamma \in [2, 10]$ dB. 

As shown in Fig. \ref{fig:fig_adapt_appendix}, the fully-adaptive model along with the models trained at fixed $SNR_{sr}$ values are tested under $\gamma \in \{4, 6\}$ dB. It can be seen that the fully-adaptive model achieves comparable performance with the models trained with fixed $SNR_{sr}$, which verifies the effectiveness of the proposed scheme. 

\figadaptappendix

\section{Conclusion}
\label{sec:conclusion}
We investigated the cooperative image transmission problem over a wireless relay channel, and proposed three different DeepJSCC schemes, namely DeepJSCC-AF, DF, and PF, all employing DNNs as encoder and decoder. All three schemes are trained in an end-to-end fashion and employ channel-attention modules so that a single trained network can be deployed across a variety of channel conditions. 
It is shown that the proposed schemes outperform separation-based protocols {employing state-of-the-art compression and channel coding schemes}, while avoiding the cliff effect. {Unlike the DF scheme for channel-coded cooperation}, the proposed DeepJSCC-DF and PF schemes are robust against poor source-to-relay channel conditions, and both schemes exhibit superior performance over DeepJSCC-AF with higher source-to-relay SNRs.

\bibliographystyle{IEEEbib}
\bibliography{refs}

\clearpage

\end{document}